\documentclass[twocolumn]{aastex63}

\usepackage{graphicx}
\usepackage{amssymb}
\usepackage{amsmath}
\usepackage{latexsym}
\usepackage{color}
\usepackage{algorithm}
\usepackage{xspace}
\usepackage{textgreek}
\usepackage[T1]{fontenc}
\usepackage{hyperref}
\usepackage[all]{hypcap}    
\hypersetup{
	colorlinks,
		citecolor=blue,
		filecolor=blue,
		linkcolor=blue,
		urlcolor=blue
	}
\newcommand{\ENZO}{\textsc{ENZO}\xspace}

\newcommand{\Romulus}{\textsc{Romulus}\xspace}
\newcommand{\RomulusC}{\textsc{RomulusC}\xspace}
\newcommand{\CHANGA}{\textsc{ChaNGa}\xspace}
\newcommand{\Ramses}{\textsc{Ramses}\xspace}
\newcommand{\RhapsodyG}{\textsc{Rhapsody-G}\xspace}

\DeclareMathSymbol{\leq}{\mathrel}{symbols}{20}
   
\DeclareMathSymbol{\geq}{\mathrel}{symbols}{21}

\shorttitle{Tests of AGN Feedback Kernels}
\shortauthors{Glines et al.}

\begin{document}
\title{Tests of AGN Feedback Kernels in Simulated Galaxy Clusters}

\author[0000-0002-6837-8195]{Forrest W. Glines}
    \email{glinesfo@msu.edu}
\affiliation{Department of Physics and Astronomy,
    Michigan State University, East Lansing, MI 48824, USA} 
\affiliation{Department of Computational Mathematics, Science, and Engineering,
    Michigan State University, East Lansing, MI 48824, USA} 
\author[0000-0002-2786-0348]{Brian W. O'Shea}
\affiliation{Department of Physics and Astronomy,
    Michigan State University, East Lansing, MI 48824, USA} 
\affiliation{Department of Computational Mathematics, Science, and Engineering,
    Michigan State University, East Lansing, MI 48824, USA} 
\affiliation{National Superconducting Cyclotron Laboratory,
    Michigan State University, East Lansing, MI 48824, USA}
\author[0000-0002-3514-0383]{G. Mark Voit}
\affiliation{Department of Physics and Astronomy,
    Michigan State University, East Lansing, MI 48824, USA} 

\label{firstpage}

\begin{abstract} 
In cool-core galaxy clusters with central cooling times much shorter than a Hubble time, condensation of the ambient central gas is regulated by a heating mechanism, probably an active galactic nucleus (AGN). Previous analytical work has suggested that certain radial distributions of heat input may result in convergence to a quasi-steady global state that does not substantively change on the timescale for radiative cooling, even if the heating and cooling are not locally in balance.  To test this hypothesis, we simulate idealized galaxy cluster halos using the \ENZO code with an idealized, spherically symmetric heat-input kernel intended to emulate.  Thermal energy is distributed with radius according to a range of kernels, in which total heating is updated to match total cooling every $10 ~\text{Myr}$. Some heating kernels can maintain quasi-steady global configurations, but no kernel we tested produces a quasi-steady state with central entropy as low as those observed in cool-core clusters. The general behavior of the simulations depends on the proportion of heating in the inner $10 ~\text{kpc}$, with low central heating leading to central cooling catastrophes, high central heating creating a central convective zone with an inverted entropy gradient, and intermediate central heating resulting in a flat central entropy profile that exceeds observations. The timescale on which our simulated halos fall into an unsteady multiphase state is proportional to the square of the cooling time of the lowest entropy gas, allowing more centrally concentrated heating to maintain a longer lasting steady state.
\end{abstract}


\section{Introduction}
\label{sec:introduction}

Cool-core (CC) clusters have X-ray surface brightness profiles with sharp central peaks produced by substantial radiative losses of thermal energy from gas within the central few tens of kpc \citep{fabian_cooling_1994}. Given the observed rates of energy loss, CC clusters should be capable of radiating away their central thermal energy in less than $1 ~\text{Gyr}$.  If uncompensated, such a rapid cooling rate would lead to a cooling catastrophe in which multiphase condensation of ambient gas into cold clouds fuels star formation rates much greater than those observed. However, CC clusters are generally not observed to experience such dramatic cooling catastrophes \citep{mcdonald_anatomy_2019}.   They apparently remain close to thermal balance for billions of years and are common, representing about half of all galaxy clusters at the present time.  Consequently, some mechanism must be counteracting central radiative cooling, and active galactic nuclei (AGN) are currently believed to be the responsible energy sources \citep{fabian_chandra_2000,mcnamara_chandra_2000,fabian_very_2006,mcnamara_heating_2007,panagoulia_volume-limited_2014,gaspari_self-regulated_2016}.

Many other heat sources have been explored, including 
galaxy cluster mergers \citep{roettiger_numerical_1997,fixed-gomez_cooling_2002,zuhone_stirring_2010}, 
supernovae \citep{ciotti_cooling_1997,wu_effect_1998,voit_regulation_2001,domainko_feedback_2004,short_heating_2013}, 
thermal conduction \citep{chandran_thermal_1998,narayan_thermal_2001,malyshkin_transport_2001,voigt_conduction_2002,jubelgas_thermal_2004,fixed-bruggen_equilibrium_2003,2013ApJ...778..152S}, 
gravitational heating \citep{khosroshahi_old_2004,dekel_gravitational_2007,dekel_gravitational_2008}, 
and gas sloshing \citep{ritchie_hydrodynamic_2002,markevitch_nonhydrostatic_2001,zuhone_stirring_2010}. 
Most either do not provide enough heat to offset the observed cooling or do not adjust to the radiative cooling rate on a short enough time scale.  Core cooling times in many CC clusters 
are $< 1 ~\text{Gyr}$ \citep{cavagnolo_intracluster_2009,fixed-pratt_galaxy_2009}, much less than the lifetimes of these clusters,
suggesting that any heating mechanism coupled to cooling must react on shorter timescales.  The gas accretion rate onto the central supermassive black hole (SMBH) would therefore need to couple to the radiative cooling rate with a lag time no greater than several hundred Myr.

Feedback from the central galaxy and AGN was explored numerically as early as \citet{tabor_elliptical_1993}, \citet{metzler_simulation_1994}, and \citet{binney_evolving_1995}.  More recently, \citet{sijacki_unified_2007}, \citet{gaspari_dance_2011}, \citet{li_cooling_2015}, \citet{meece_triggering_2017}, \citet{prasad_cool_2015,prasad_agn_2017,prasad_cool-core_2018}, and many others \citep{fabjan_simulating_2010,dubois_jet-regulated_2010,short_heating_2013,yang_how_2016} 
have demonstrated in hydrodynamic simulations of idealized galaxy clusters that AGN can plausibly regulate the high cooling rate in CC clusters.  Simulated AGN self-regulate by coupling feedback energy output to the ambient gas density or cold-gas accretion rate around the AGN and inject that energy through either thermal deposition around the AGN or bipolar outflows from the AGN or a combination of the two.
In addition to regulating the cooling rate and the condensation of cold gas clouds within the cluster, some of these AGN simulations produce temperature, density, and entropy profiles that resemble observations, including the multiphase cores observed in the central $100 ~\text{kpc}$ of galaxy clusters \citep{gaspari_cause_2012,meece_triggering_2017,prasad_cool-core_2018}.

The simulations that most successfully resemble observations rely on cold-gas accretion to fuel the AGN and bipolar outflows to distribute the feedback energy \citep{gaspari_raining_2017,fixed-gaspari_unifying_2017,voit_global_2017,meece_triggering_2017}. Ambient gas at the center of the system is nearly isentropic and therefore convectively unstable, resulting in the formation of a complex multiphase medium in which cold clumps of gas condense out of the ambient gas and precipitate onto the black hole.  As the precipitation increases, so does the output of feedback energy, which raises the central cooling time and ultimately reduces the rate of precipitation.  The resulting coupling suspends the ambient medium in a transitional state on the verge of a cooling catastrophe.  Condensation outside of the isentropic center is marginally suppressed by buoyancy, and gas lifted out of the center by bipolar jets and buoyant bubbles forms multiphase filaments
\citep{fixed-revaz_formation_2008,li_modeling_2014,li_modeling_2014-1}, in general agreement with observations \citep{mcdonald_origin_2010,russell_alma_2016,russell_alma_2017}.
However, even these idealized simulations do not track all of the physical processes that might be transporting and thermalizing AGN feedback energy, which range from turbulent heat diffusion \citep{ruszkowski_galaxy_2011,zhuravleva_turbulent_2014}, viscous dissipation of waves generated by the AGN \citep{fixed-ruszkowski_cluster_2004}, and cosmic rays created by the AGN heating the plasma via small scale fluid instabilities \citep{fixed-boehringer_dynamical_1988,loewenstein_cosmic-ray_1991,rephaeli_energetic_1995,fixed-colafrancesco_cooling_2004,fixed-pfrommer_simulating_2007,fixed-jubelgas_cosmic_2008}.

Incorporating all of these mechanisms and processes into a cosmological simulation of galaxy cluster formation is currently prohibitively complex.  Typically, the minimum spatial resolution in simulations modeling hot jets that interact with the intracluster medium is $200 ~\text{pc}$. The finer resolution of the gas along which the jet deposits energy leads the jet to drill a hole through the ICM, allowing energy from the AGN to be deposited at further radii \citep{meece_triggering_2017,li_cooling_2015}.  These resolution constraints are not always feasible for large cosmological simulations, because the computational effort needed to model these AGN jets exerts unacceptable drag on the evolution of the entire system.  
Therefore, simplified subgrid models are still needed to represent AGN feedback in cosmological simulations.

The results we present here emerged from an effort to develop a simple 
heat-input kernel to serve as an acceptable proxy for the much more complex process of AGN feedback.  We sought a kernel that would satisfy three criteria:
\begin{enumerate}
    \item The simulated hot-gas atmospheres of clusters balanced by AGN feedback should remain nearly thermally steady, meaning that they should not dramatically change because of cooling and feedback for periods of several billion years.
    \item The central entropy of the hot gas in such a quasi-steady cluster halo should not exceed the values observed in CC clusters.
    \item The feedback process should be computationally efficient, requiring neither very high resolution nor extremely small time steps that would make implementation in a current cosmological simulation prohibitively costly.
\end{enumerate}

The first criterion requires the heating kernel to prevent a cooling catastrophe, which we define for the purposes of this paper to be a factor of 10 increase in radiative cooling within $10~\text{Myr}$, accompanied by a rapid increase in the amount of cold ($10^4~\text{K}$) gas. As the central cooling time becomes short, compensating thermal feedback is needed to prevent runaway overcooling.

The second criterion requires that the kernel not overheat the central region, which would elevate or invert the central entropy profile. Such centrally concentrated AGN feedback can produce both non-cool core (NCC) clusters or observationally unreasonable galaxy clusters with large central entropy peaks. 
Furthermore, buoyancy is unable to suppress runaway thermal instabilities in systems with centrally flattened entropy profiles, making them prone to multiphase condensation \citep[e.g.,][]{voit_global_2017} Simultaneously satisfying both this criterion and the first one proved to be difficult, even though observations show that CC clusters can remain remarkably close to a cooling catastrophe without producing an overabundance of cold gas and young stars.

Finding a way to satisfy the third criterion along with with the other two was the main motivator for this paper.  Tracking the rapid formation of a complex multiphase medium approaching a cooling catastrophe requires high resolution and small time steps. Furthermore, if feedback energy output is directly linked to condensation of cold clouds, the approach of a cooling catastrophe leads directly to rapid central heating, 
thereby compounding the computational requirements.  We therefore sought a simple method that would avert a cooling catastrophe while still allowing the ambient central gas to remain in a low-entropy state.

In our search for a numerically simple heating kernel that would satisfy these three criteria, we investigated kernels with a power-law radial distribution of thermal feedback, normalized so that feedback heating globally equals radiative cooling within the galaxy-cluster halo. Use of such a heating kernel implicitly assumes that the most consequential feature of more complex AGN feedback mechanisms is the radial distribution of heat input. Depositing heat into the gas according to a kernel that depends only on radius is numerically simple and efficient to incorporate into cosmological simulations, and it does not require high spatial resolution as long as the feedback method can maintain the hot halo gas in a thermally steady state without overcooling. In order to create a tunable model, we also modified the radial power law with an inner truncation radius to limit central feedback and an outer exponential cutoff radius to constrain the bulk of the AGN heating to gas with shorter and more relevant cooling times. These additional parameters gave us a numerically simple but tunable model to search for an adequate AGN feedback kernel. We heuristically explored different values of the inner truncation radius that avoided central entropy peaks and different values of the outer cutoff radius that kept the majority of the feedback inside the region of the halo where gas cools within a hubble time. We discuss the model in more detail later in the paper.

Section \ref{sec:methodology} discuses the simulation setup and AGN feedback prescription and heating kernel in detail. 
Section \ref{sec:results} shows simulation results,  describing in detail the results of three heating kernels that broadly represent the whole set of simulations, and examining the impact of different heating kernel parameters.
Section \ref{sec:discussion} discusses the adequacy of the heating kernels tested, the robustness of the resulting feedback model, and the possible implications of these simulations for our understanding of AGN feedback in general. 
Lastly, Section \ref{sec:summary} summarizes the results and conclusions of this work.

\section{Methodology}
\label{sec:methodology}

This work builds upon simulations by \cite{meece_triggering_2017}, using the same initial conditions  from that work, described in \S \ref{sec:simulation_setup}, but using an AGN feedback kernel that is adapted to deposit energy at large radii as described in \S \ref{sec:AGN_feedback}.

\subsection{Simulation Setup}
\label{sec:simulation_setup}

We ran several simulations of idealized galaxy cluster halos with a simplified AGN
heating model using the hydrodynamics code \ENZO \citep{bryan_enzo:_2014}.

We used initial conditions approximating 
the Perseus Cluster, following the approach from \citet{li_simulating_2012} and
\citet{meece_triggering_2017}. The ICM begins as a hydrostatic sphere of gas in
a fixed gravitational potential. 

The gravitational potential has two components: a dark matter halo
profile and a BCG with a mass profile with parameters chosen to match the Perseus
cluster. 
The dark matter follows the NFW profile \citep{NFW_1997},
using $M_{200c} = 8.5 \times 10^{14} \text{M}_\odot$ for the mass within the virial radius and a concentration parameter $c
= 6.81$.  The dark matter density from the NFW profile takes the form
\begin{equation}
  \rho^{\text{NFW}}(r) = \frac{ \rho_0^{\text{NFW}} }
  {\left ( r / R_s \right )
  \left( 1 + \frac{r}{R_s} \right)^2}
\end{equation}
where the scale density $\rho_0^{\text{NFW}}$ is defined by 
\begin{equation}
    \rho_0^{\text{NFW}} = \frac{ 200}{3} \rho_c \frac{c^3}{ \ln(1 + c) - c/\left ( 1+c\right )},
\end{equation}
where $\rho_c=3 H^2/\left ( 8 \pi G \right )$ is the critical density and the scale radius $R_s$ can be found from
\begin{equation}
    M_{200c} = 4 \pi \rho_0^{\text{NFW}} R_s^3 \left [ \ln \left ( 1 + c \right ) - c/\left ( 1+c\right ) \right ].
\end{equation}

The BCG mass profile, following \citet{meece_triggering_2017}, has the
form
\begin{equation}
  M_*(r) = M_4 \left [ \frac{ 2^{-\beta_*} }{
    \left ( r/4 \text{ kpc} \right ) ^{-\alpha_*} 
    \left ( 1 + r/ 4 \text{ kpc} \right )^{\alpha_* - \beta_*} } \right ],
\end{equation}
where $M_4 = 7.5 \times 10^{10} M_\odot$ is the stellar mass within $4 \text{kpc}$ and $\alpha_* =
0.1$ and $\beta_* = 1.43$ are constraints.\footnote{Due to a programming error, the simulations use an incorrect initial mass profile for the BCG, which leads to the central $1 ~\text{kpc}$ being initialized out of hydrostatic equilibrium, with an absence of baryonic mass by less than a factor of two.  However, the central halo gas either relaxes to hydrostatic equilibrium within $50 ~\text{Myr}$ or AGN feedback quickly drives it further from equilibrium, depending on the heating kernel parameters. Consequently, this error in the initial conditions} does not substantially affect our results.

The initial pressure was computed from
the temperature and density assuming an ideal gas with $\gamma=5/3$ in hydrostatic equilibrium with the gravitational potential. 
Cosmological expansion is neglected in these simulations.  We
used a vanilla {\textLambda}CDM model to get the virial mass of the NFW halo and to
set its gas temperature. We set redshift $z=0$ at initialization with
$\Omega_M = 0.3$, $\Omega_\Lambda = 0.7$, and $H_0 = 70 \text{ km s}^{-1}$.  We note that the precise details of the cosmological model do not impact the results presented in later sections of this paper, which pertain to baryonic physics in the halo core.

The entropy profile of the gas, using the form
\begin{equation}
  K \equiv \frac{ k_bT}{n_e^{2/3} }
\end{equation}
for the specific entropy,
where $k_b$ is Boltzmann's constant, $T$ is the temperature, and $n_e$ is the electron density,
was initialized to a power law
\begin{equation}
  K(r) = K_{0} + K_{100} \left ( r/ 100 \text{ kpc} \right )^{\alpha_K},
\end{equation}
following the power law fits used in the ACCEPT database
\citep{cavagnolo_intracluster_2009}. Here, $r$ is the radius from the halo
center and $K_0=19.38 ~\text{keV cm}^2 , K_{100}=119.87 ~\text{keV cm}^2$, and $\alpha_K=1.74$ are fitting
parameters corresponding to the core entropy, entropy slope and exponential
increase, chosen to approximate the Perseus Cluster.

The simulations were run on a cartesian grid in a cubic volume with side length of $3.2~\text{Mpc}$, with $64^3$ cells in the
base grid of the AMR hierarchy and a maximum of $8$ levels of refinement,
making the resolution of the finest cells approximately $195~\text{pc}$. The mesh was
refined based on the magnitude of gradients in fluid quantities and high baryon density. Additionally, a cubic grid with side length
$4~\text{kpc}$ around the simulation center and was fixed at the maximum level of
refinement with $195~\text{pc}$ resolution.  

Each simulation was allowed to run for $16~\text{Gyr}$ or until excessive AGN feedback during a cooling catastrophe either created unphysical cell values or led to intractably small timesteps (see Section \ref{sec:robustness_of_heating_prescription}).  
To give context to the simulation duration, consider that the sound speed of gas with a temperature of $T=2 \times 10^7 \text{ K}$ is $c_s = \sqrt{\gamma k_B T/ \mu m_{\rm H}} \approx 0.70 \text{ Mpc} \text{  Gyr}^{-1}$, where $m_{\rm H}$ is the mass of hydrogen and $\mu=0.6$ is the mean mass per particle in units of $m_{\rm H}$, meaning that
the approximate sound crossing time across the inner $R = 0.5 \text{ Mpc}$, where the majority of the dynamics of the galaxy cluster halo evolves, is approximately 1.4~Gyr.

We used the ZEUS solver for hydrodynamics \citep{stone_zeus-2d:_1992} due to
its robustness to evolve through discontinuities in the fluid around the AGN
due to sharply peaked thermal injection. ZEUS is a relatively diffusive solver
and requires an artificial viscosity, which may affect the accuracy of the
hydrodynamics simulation \citep{stone_zeus-2d:_1992,meece_jr_agn_2016}.
Tabulated cooling was used to model radiative cooling following
\citet{schure_new_2009}, assuming a metallicity of 0.5 Z$_\odot$.
The cooling table has a temperature floor of $10^4~\text{K}$; any processes
below this temperature will take place on a smaller scale than can be accurately explored with our spatial
resolution.   

Simulation results were analyzed using \texttt{yt} \citep{turk_yt:_2011}.

\subsection{AGN Feedback Kernels}
\label{sec:AGN_feedback}

\begin{figure}
  \begin{center}
    \includegraphics[width=1\linewidth]{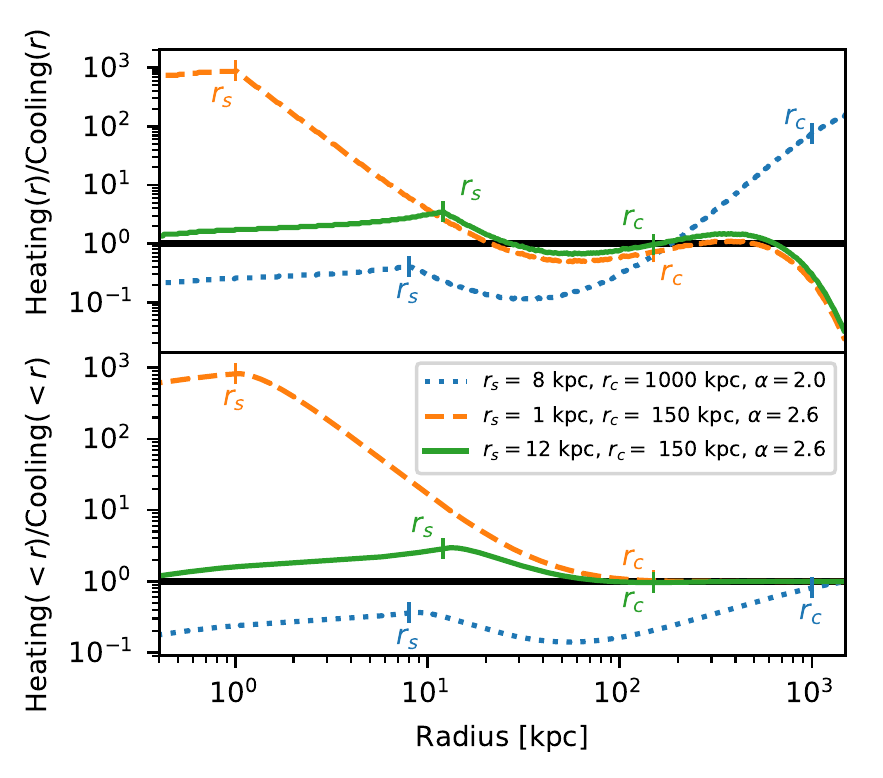}
  \end{center}
  \caption{
    \label{fig:heating_ratio}
    \textbf{Top:} Local ratio of heating to cooling as a function of radius ($r$) at the beginning of several representative simulations. The dotted blue line shows a simulation with low central heating and heating kernel parameters $\alpha=2.0$, $r_s= 8 ~\text{kpc}$, and $r_c = 1000 ~\text{kpc}$. The dashed orange line shows a simulation with high central heating and heating kernel parameters $\alpha=2.6$, $r_s=1 ~\text{kpc}$, and $r_c=150 ~\text{kpc}$. The solid green line shows a simulation with intermediate central heating and heating kernel parameters $\alpha=2.6$, $r_s=12 ~\text{kpc}$, and $r_c=150 ~\text{kpc}$.
\textbf{Bottom:} Cumulative ratio of heating to cooling within $r$ for the same simulations. At large radii, all of the cumulative heating curves converge to the cumulative cooling rate because total heating is normalized to equal to total cooling rate at $R=1.5 ~\text{Mpc}$.  
  }
 \end{figure}

In our simplified AGN feedback model, thermal energy is deposited in a spherically symmetric distribution around the halo center by an assumed AGN, with
the total amount of heating set equal to the total cooling in the halo every $10~\text{Myr}$.  Heating per unit volume $\dot e(r)$ is distributed following a power law in radius so that $\dot e (r) \propto r^{-\alpha}$.  This basic power-law functional form has several numerical and practical issues.  Most critically, these issues are a volumetric heating rate that diverges to infinity at the 
halo center,  a ``long tail'' of heating at the halo outskirts where cooling is too slow to be relevant, and an unrealistic hard cutoff at the simulation boundaries.  These latter two issues are compounded by observations that suggest AGN feedback is generally constrained to be within a few hundred kpc of the halo center.  To address these issues and to create a more tunable and effective heating kernel, we added two parameters: a minimum truncation radius $r_s$ (effectively a smoothing length) and an exponential decay cutoff radius $r_c$. To avoid having the feedback stop at a simulation boundary at $x,y,z=\pm 1.6 ~\text{Mpc}$, the AGN feedback is contained within a radius of $R=1.5~\text{Mpc}$ and set to zero outside this radius. Since the heating leading up to $R$ is negligible compared to the cooling at far radii and the cooling time of the gas is much longer than the simulation time at that radius, we do not expect the value of $R$ to have an impact on the outcome of the simulation.
The full form of the feedback kernel defining the heating rate per unit volume $\dot e (t) [\text{erg} \text{ s}^{-1} \text{ cm}^{-3}]$ is
\begin{equation}
  \dot{e}(r,t) = \frac{\dot{E}(t)}{A}
  \left\{\begin{matrix}
    \left( \frac{r_s}{r_c}\right )^{-\alpha} \exp {\left ( -\frac{r_s}{r_c} \right )},
    &r \leq r_s \\
    \left( \frac{r}{r_c} \right )^{-\alpha} \exp { \left ( -\frac{r}{r_c} \right )},
    &r_s <r \leq R \\
      0, & R < r
  \end{matrix}\right. .
\end{equation}
The scalar $A ~\text{[cm}^{3}\text{]}$ is defined by
\begin{eqnarray}
  A & = & \int_0^{r_s }
    4 \pi r^2 dr
    \left( \frac{r_s}{r_c} \right )^{-\alpha} 
    \exp {\left ( -\frac{r_s}{r_c} \right )} 
    \nonumber \\
    &  & + \, \int_{r_s }^R
     4 \pi r^2 dr
    \left( \frac{r}{r_c} \right )^{-\alpha} \exp { \left ( -\frac{r}{r_c} \right )} \\
    & = & \frac{4 \pi}{3} 
    \exp{ \left ( -\frac{r_s}{r_c} \right )} r_s^3 
        \left ( \frac{r_s }{r_c} \right )^{-\alpha} 
        \nonumber \\
    & & + \, 4 \pi r_c^3 \left [ 
      -\Gamma \left ( 3 - \alpha, \frac{R}{r_c} \right ) 
      -\Gamma \left ( 3 - \alpha, \frac{r_s}{r_c} \right ) 
                     \right ],
\end{eqnarray}
where $\Gamma(s,x) = \int_x^\infty t^{s-1} e^{-t} dt$ is the upper incomplete
gamma function, normalizes $\dot{e} (t)$ so that the integral of $\dot{e} (t)$ over the volume of the simulation matches $\dot{E}(t)$.  Higher values of $\alpha$ correspond to more centralized feedback around the AGN. Without the inner smoothing length, a heating kernel with $\alpha \geq 3$ is not normalizable, because integration over a volume containing the origin diverges.

The total heating rate $\dot E(t)$ is set to the total cooling rate within the cluster halo. Since the total cooling rate can be difficult to compute on-the-fly due to
the nature of the AMR hierarchy's timestep update, it is recomputed only every $10~\text{Myr}$. Although the cooling rate increases exponentially leading up to a cooling catastrophe, the
increase is slow enough that the heating rate does not fall behind the true cooling rate by more than a few percent except immediately within a Myr before the catastrophe, at which point the simulation has already demonstrated that the particular heating kernel being tested is inadequate.

Note that the short time scale over which heating reacts to cooling in our model is not physical. Heat deposition resulting from AGN feedback does not instantaneously happen far from the AGN.  We therefore probed heating kernels with a $50 ~\text{Myr}$ lag time between heating and cooling as well as averaging cooling over the same time period to smooth out jumps in heating. However, adding lag time led to more cold gas forming due to the lack of immediate feedback to counter condensation and more explosive feedback overall.

This study tested 91 different heating kernels with a range of parameters: different radial exponents $\alpha\in [2.0,3.2]$, smoothing lengths $r_s\in{1,4,8,10,12,16,20,40} ~\text{kpc}$, and exponential cutoff radii $r_c \in {100,150,200} ~\text{kpc}$. We began our exploration of the parameter space by setting $r_s=1 ~\text{kpc}$ and $r_c=1500 ~\text{kpc}$ and sampled the range of $\alpha$ before trying different values of $r_s$ and $r_c$ with a smaller number of $\alpha$ values, seeking parameter combinations that seemed closest to an optimal kernel.
Figure \ref{fig:heating_ratio} presents a representative sampling of heating kernels showing the initial ratio of heating to cooling as a function of radius, including both the local ratio at each radius and the cumulative ratio within each radius.  Table \ref{tab:paramters} lists all combinations of parameters explored.

\begin{table}
\caption{
\label{tab:paramters} List of combinations of inner smoothing radius $r_s ~\text{[kpc]}$, outer cutoff radius $r_c ~\text{[kpc]}$, and exponent $\alpha$ used. The rightmost column lists all values of $\alpha$ explored for the given combination of $r_s$ and $r_c$ in the leftmost and middle column.
}
\begin{center}
\begin{tabular}{ | l | l | p{50mm}| }
\hline
$r_s$ [kpc] & $ r_c$ [kpc] & $\alpha$ \\
\hline \hline
1 & 150 & 2.0, 2.2, 2.4, 2.6, 2.8, 3.0, 3.2 \\
\hline
1 & 1000 & 2.0, 2.1, 2.2, 2.3, 2.35, 2.375, 2.4, 2.425, 2.45, 2.5, 2.525, 2.55, 2.575, 2.6, 2.65, 2.7, 2.8, 2.9, 3.0\\
\hline
4 & 150 & 2.0, 2.2, 2.4, 2.6, 2.8, 3.0, 3.2 \\
\hline
8 & 150 & 2.0, 2.2, 2.4, 2.6, 2.8, 2.9, 2.95, 3.0, 3.2 \\
\hline
8 & 1000 & 2.0, 2.2, 2.4, 2.6, 2.8, 3.0, 3.2 \\
\hline
16 & 150 & 2.0, 2.2, 2.4, 2.6, 2.8, 3.0, 3.2 \\
\hline
10 & 150 & 2.0, 2.2, 2.4, 2.6, 2.8, 3.0, 3.2 \\
\hline
10 & 150 & 2.0, 2.2, 2.4, 2.6, 2.8, 3.0, 3.2 \\
\hline
12 & 150 & 2.0, 2.2, 2.4, 2.6, 2.8, 3.0, 3.2 \\
\hline
16 & 100 & 2.0, 2.2, 2.4, 2.6, 2.8, 3.0, 3.2 \\
\hline
16 & 150 & 2.0, 2.2, 2.4, 2.6, 2.8, 3.0, 3.2 \\
\hline
20 & 100 & 2.0, 2.2, 2.4, 2.6, 2.8, 3.0, 3.2 \\
\hline
40 & 150 & 2.0, 2.2, 2.4, 2.6, 2.8, 3.0, 3.2 \\
\hline
\end{tabular}
\end{center}
\end{table}


\section{Results}
\label{sec:results}

\begin{figure*}
  \begin{center}
    \includegraphics[width=190mm]{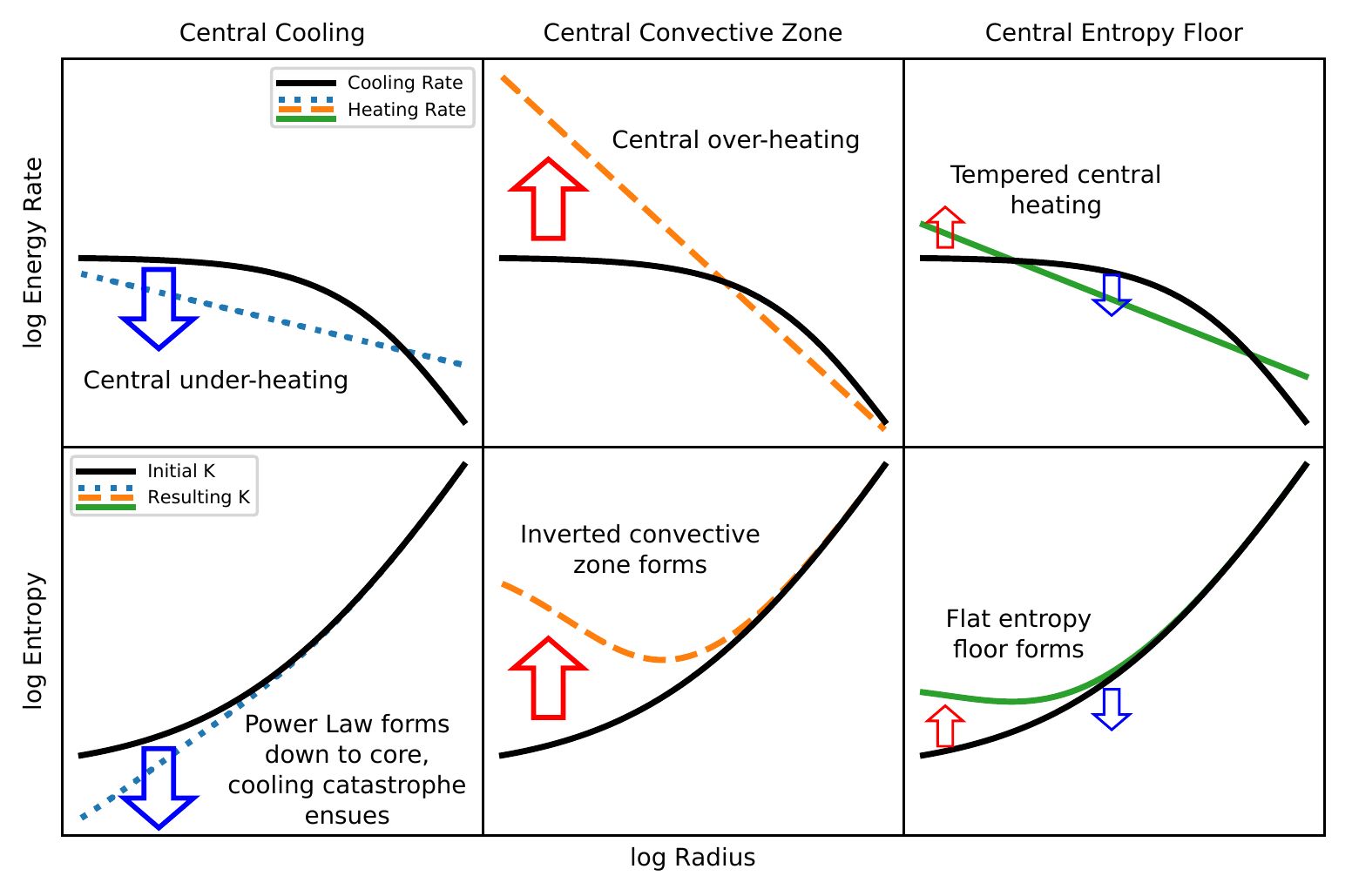}
  \end{center}
  \caption{
    \label{fig:hc_schematic}
    Schematic illustrations of how different AGN heating kernels affect the entropy profile of a simulated galaxy cluster.  In each case, the total heating rate is set equal to the total cooling rate.  \textbf{Top:} Radial profiles of radiative cooling and AGN heating per unit volume, with the initial median cooling rate in black and the AGN heating kernel in color. \textbf{Bottom:}  Response of the median entropy profile to heat input.  The initial median profile in black and the response is in color.  The left column shows a heating kernel with central heating that falls below central cooling.  The entropy profile in this case tends to follow a power law down to the origin and eventually leads to a central cooling catastrophe. The center column shows a heating kernel with excessive central heating, which elevates central entropy, inverts the entropy profile, and produces a central convective zone. The right column shows a heating kernel with intermediate central heating, which slightly raises the central entropy and produces a flat core. Due to the high initial entropy and long cooling time at outer radii, the power-law at the outer radii changes very slowly with under- and over-heating.
  }
 \end{figure*}
 
All the heating kernels we explored resulted either in cooling catastrophes within a few Gyr, central entropy levels greater than observations, or both.
Simulations that eventually formed cold, condensed gas all went through cooling catastrophes. 
In those simulations, the minimum entropy drops over time,
eventually leading to multiphase condensation. As cold clumps of gas form and runaway cooling begins, the requirement for total heating to match total cooling causes the heating rate to spike. 
The time required for cold gas to form is roughly correlated with the smallest radius at which cooling exceeds heating.  If central cooling exceeds central heating, the halo quickly forms cold gas and experiences a cooling catastrophe.
Simulations with higher central heating tend to have high central entropy, similar to observations of NCC clusters. If the heating exceeds cooling out to radii of several tens of kpc, then the simulations persist for many Gyr without forming cold gas. Under- and over-heating at outer radii beyond ~$100~\text{kpc}$ is inconsequential since the time scale of heating is much longer than the dynamical time scale of the system due to the large specific energy and entropy at initialization.

Figure~\ref{fig:hc_schematic} schematically shows the general behavior of the different heating kernels.  The three heating kernel examples in Figure~\ref{fig:heating_ratio} have colors that match the corresponding schematics in Figure~\ref{fig:hc_schematic}.
Figure~\ref{fig:phasePlots} shows mass density profiles of cooling rate, heating rate, and entropy at later moments in simulations employing the same three heating kernels as in Figure~\ref{fig:heating_ratio}.

\subsection{Categorization of Simulations}
\label{sec:categorization_of_simulations}

The results of our simulations can be grouped according to the morphology of the entropy profiles that develop within the central $100 ~\text{kpc}$:
\begin{enumerate}

\item \textbf{Central Cooling.}  The entropy profiles of simulated cluster halos with heating that is insufficient to balance radiative cooling at small radii develop central cooling flows with a positive entropy gradient at all radii.  They undergo a central cooling catastrophe relatively quickly, in which runaway multiphase condensation at small radii brings the simulation to a halt.  
\item  \textbf{Central Convective Zone.}  The entropy profiles of simulations with high central heating form an inner convective zone with high central entropy and a negative central entropy gradient.  Those simulations persist the longest before undergoing cooling catastrophes.  
\item \textbf{Central Entropy Floor.} Simulations with intermediate central heating can maintain a nearly flat entropy gradient within the central $\sim 10$ to 20~kpc. 
\end{enumerate}
For the purposes of our analysis, we define these categories based on the entropy within the inner $25 ~\text{kpc}$. We categorize as Central Cooling those simulations whose average minimum entropy remains below $12 ~\text{keV cm}^2$ (2/3 of the the initial minimum central entropy of  $18 ~\text{keV cm}^2$) .  The Central Convective Zone simulations are defined to have maximum central entropy above $50 ~\text{keV cm}^2$ (equal to the initial mean entropy of the inner $100 ~\text{kpc}$).  No simulation meets both of these criteria, so there is no overlap of these first two groups.  The remaining simulations, which have minimum central entropies above $12 ~\text{keV cm}^2$ and maximum central entropies below $50 ~\text{keV cm}^2$, are categorized as Central Entropy Floor simulations. 

\begin{figure*}
	\begin{center}
    \includegraphics[width=190mm]{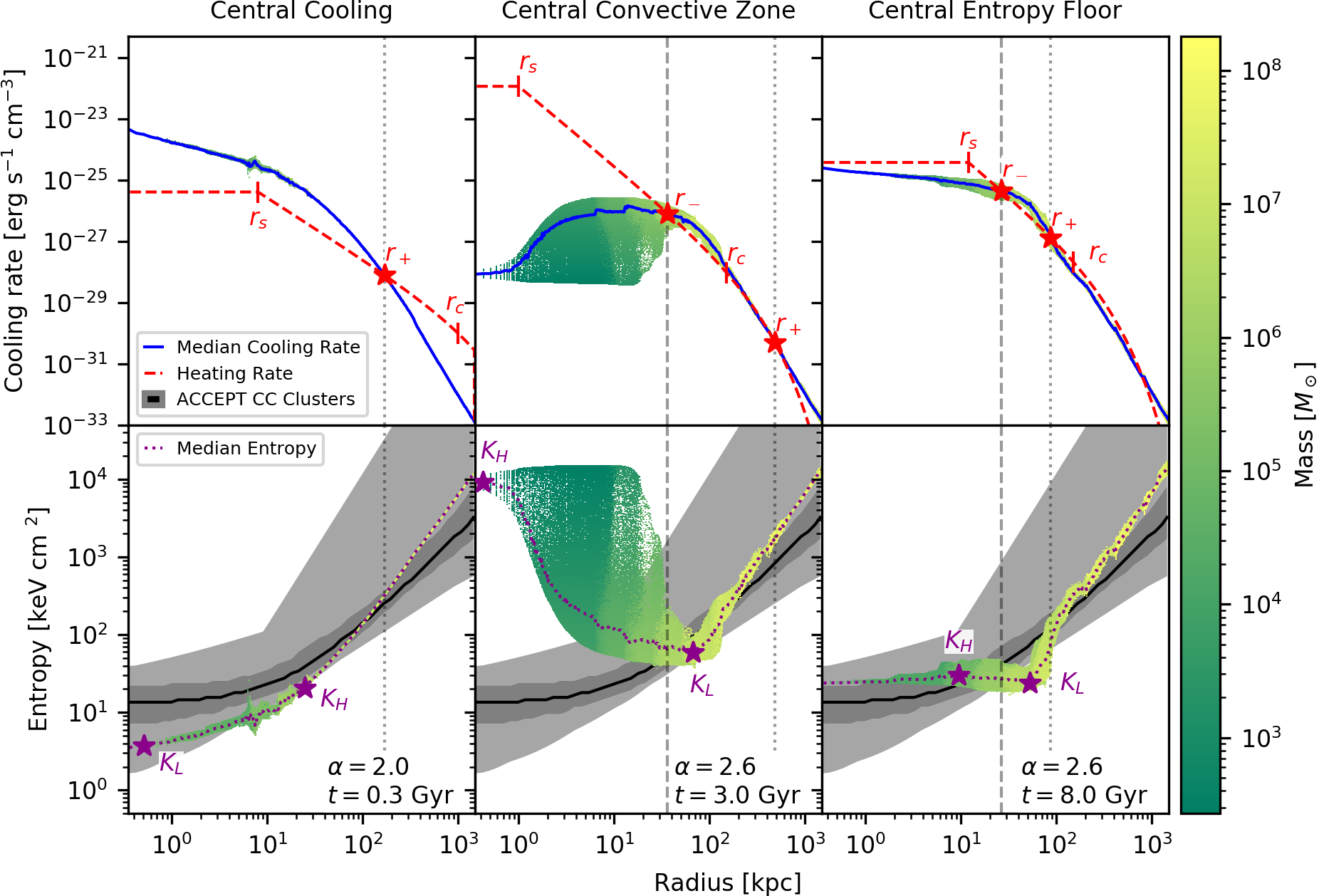}
	\end{center}
	\caption{
    \label{fig:phasePlots}
    Mass density plots of cooling and heating rate (\textbf{top}) and entropy  (\textbf{bottom}) versus radius, with color representing the total mass of all simulation cells from a 2D histogram of cooling rate and entropy versus radius. Across the three columns we show three simulations at different times that broadly represent the whole set of simulations, as differentiated by the behavior of the inner tens of kpc. The left column shows a simulation (with $\alpha = 2.0$, $r_s = 8 ~\text{kpc}$, and $r_c = 1000 ~\text{kpc}$ at $t=0.3$ Gyr) with low central heating which allows excess central cooling that quickly undergoes a cooling catastrophe. The middle column shows a simulation (with $\alpha = 2.6$, $r_s = 1 ~\text{kpc}$, and $r_c = 150 ~\text{kpc}$ at $t=3.0$ Gyr) with high central heating that maintains a convective zone in the inner $100 ~\text{kpc}$ with a high central entropy peak. The right column shows a simulation (with $\alpha = 2.6$, $r_s = 12 ~\text{kpc}$, and $r_c = 150 ~\text{kpc}$ at $t=8.0$ Gyr) with an intermediate amount of central heating and that holds a flat entropy floor slightly elevated from the initial conditions and observational data on the entropy of the inner tens of kpc. On the entropy plots, observational entropy data of clusters from the ACCEPT data set are displayed in grayscale showing the range (light grey), 68\% confidence interval (dark grey), and median (black line) of the dataset. The median entropy is also marked by a magenta line, and the minimum ($K_L$) and maximum ($K_H$) values of the entropy median within the inner $25 ~\text{kpc}$ are marked by stars. On the cooling rate plots, the heating rate is marked by a red line and the median cooling rate is marked by a blue line. The crossover radii $r_-$ and $r_+$ as defined in the text are marked by stars in the simulations where they can be defined.The heating curve parameters $r_s$ and $r_c$ are also annotated with finely dashed and dashed gray lines.
  }
\end{figure*}

The schematic diagrams in Figure \ref{fig:hc_schematic} illustrate the general behavior of the different categories.  Figure \ref{fig:phasePlots} shows representative snapshots of both cooling rate and entropy versus radius.  Some of our simulations exhibit behavior from multiple categories at different times in their evolution.  The following subsections describe each category in more detail.  

\subsubsection{Central Cooling}

Simulations with low $\alpha$, large $r_c$, or large $r_s$ tend to have central cooling exceeding central heating, which quickly leads to a cooling catastrophe. The left column in Fig. \ref{fig:phasePlots} shows an example of such a simulation. Within the inner $10 ~\text{kpc}$, the heating rate ranges from half the cooling rate to more than an order of magnitude less than the cooling rate.  Because the central heating is insufficient to counteract a growing mass of strongly cooling gas at the halo center, the simulation produces a cooling catastrophe within $2 ~\text{Gyr}$.  However, up to the moment at which a substantial quantity of cold gas forms, the entropy profile remains close to the initial state and similar to the cool-core clusters in the ACCEPT data set.  

\subsubsection{Central Convective Zone}
Heating rates within the central $\sim 10$~kpc of simulations with high $\alpha$, small $r_c$, or small $r_s$ tend to greatly exceed radiative cooling.  The middle column in Fig. \ref{fig:phasePlots} shows an example.  Excess central heating leads to a central entropy peak and an inverted entropy profile that drives convection.  Low-entropy gas at the minimum entropy point sinks toward the center, but is reheated there and eventually rises to larger radii.  Such a convective configuration can persist for many Gyr without producing multiphase condensation, because the minimum entropy and minimum cooling time are both large.  

A few of the simulations in this category do form multiphase gas.  When that happens, condensation first appears at the minimum of the entropy profile and rapidly leads to a cooling catastrophe. Although these simulations have large central heating rates, the heating rate still falls below cooling at intermediate radii (near the entropy minimum), allowing large clumps of cold gas to form there.  In all cases in which a convective central zone forms, the central entropy is excessive compared with observed CC clusters, in some cases being more typical for a NCC.

\subsubsection{Central Entropy Floor}
Simulations with intermediate central heating, corresponding to a narrow range of combinations of $\alpha$, $r_s$, and $r_c$, are able to maintain quasi-stable flat entropy profiles out to radii exceeding 10~kpc.  The right column in Fig. \ref{fig:phasePlots} shows an example.  Central heating within the inner $10 ~\text{kpc}$ of these simulations is typically several times the central cooling rate, sufficient to offset runaway cooling but not great enough  to produce a large entropy inversion. Only some of these simulations form cold gas, and typically do so at larger radii and later times than in the Central Cooling simulations. However, the central heating in these simulations is still great enough to elevate the central entropy above the values observed in CC clusters.

\subsection{Important radii: $r_L$, $r_H$, $r_-$, $r_+$, and $r_{\text{multi}}$}
\label{sec:useful_quantities} 

To help with the analysis of the simulations, we identify several quantities that proved to be useful for interpreting their behavior. Those quantities are labeled in Figure \ref{fig:phasePlots}.

The maximum and minimum entropy levels in the central regions turn out to be closely related to the time it takes for a cooling catastrophe to manifest.  To quantify those extremes we first determine the median entropy at each radius, illustrated by the purple dotted lines in Figure \ref{fig:phasePlots}.  We then define $K_L$ to be the minimum of the median entropy profile and $r_L$ to be the radius at that point.  Outside of $r_L$ the median entropy profile is stable to convection, but inside of $r_L$ it is convectively unstable. In simulations with low central heating, $r_L$ is close to the center.  We define $K_H$ to be the maximum of the median entropy profile within $25 ~\text{kpc}$ of the simulation center and $r_H$ to be the radius at that point. We use the $25 ~\text{kpc}$ cutoff to exclude cosmologically heated gas at large radii from the analysis in order to focus on the effects of feedback heating.  The initial entropy at $25 ~\text{kpc}$ is just below $30 ~\text{keV cm}^2$, so a persistent $K_H$ above $30 ~\text{keV cm}^2$ indicates that heating has elevated the central entropy, making it too great for a CC cluster and possibly producing a central convective zone. 

The entropy extrema $K_L$ and $K_H$ and the corresponding radii $r_L$ and $r_H$ evolve over time as feedback alters the median entropy profile.   We denote the cooling times at those radii by $t_c\left(r_L\right)$ and $t_c\left(r_H\right)$.  The value of $t_c(r_L)$ is closely linked to the time required for condensation to begin.  The relationship between how the heating kernel parameters affect $K_H$ and $K_L$ along with the associated radii and cooling times is explored in sections \ref{sec:formation_of_cold_gas}, \ref{sec:central_heating}, and \ref{sec:no_adequate_heating_kernel}.
 
The radii at which heating equals cooling are special and come in two types.  For one type, the net heating rate goes from positive to negative as $r$ increases.  We define $r_-$ to be the smallest such radius.  Excess heating within that radius tends to raise the median entropy while excess cooling at large radii causes the median entropy to decline.  The result is flattening and sometimes inversion of the median entropy profile, which drives convection and ultimately makes the system prone to condensation near $r_-$.  However, if cooling dominates heating in the central regions, then $r_-$ is undefined.   
Some relationships between $r_-$ and the simulation outcomes are explored in Section \ref{sec:formation_of_cold_gas}.

At the other type of heating-cooling equality radius, the net heating rate goes from negative to positive as $r$ increases.  We define  $r_+$ to be the largest such radius.  Outside of $r_+$, net heating raises the median entropy and suppresses condensation.  Within $r_+$, net cooling lowers the median entropy.  Together, these effects produce a positive entropy gradient in the vicinity of $r_+$.   

While the median cooling rate may exceed the heating rate at very large radii (on the order of hundreds of kpc), cooling times at those radii are so long that cold gas does not form on an astrophysically significant time scale.  During a given simulation, the radii $r_-$ and $r_+$ do not stay fixed, but rather shift as heating and cooling change the median cooling rate.  We denote the cooling time at those radii as $t_c(r_-)$ and $t_c(r_+)$. 

The heating kernel parameters also affect when cold gas forms in the simulations and at what radius the cold gas first appears. We define $t_{\text{multi}}$ to be the time from the beginning of the simulation to the moment when multiphase condensation produces cold gas. In our analysis, we use $10^5 ~\text{K}$ as the temperature cutoff for cold, although gas around these temperatures will rapidly cool to colder temperatures. Our temporal resolution of $t_{\text{multi}}$ is limited by the frequency of output to disk, which is every $10 ~\text{Myr}$.  We define $r_{\text{multi}}$ to be the radius at which cold gas first appears, using the innermost radius if cold gas appears simultaneously at multiple radii. The relationship between $r_{\text{multi}}$, $r_H$, $t_{\text{multi}}$, and $t_c(r_-)$ is explored in Section \ref{sec:formation_of_cold_gas}.

Table \ref{tab:important_variables} summarizes the variables defined in this section. These variables are used in figures and analysis in later sections.

\begin{table}
\caption{
\label{tab:important_variables}
Brief definition of variables described in full in text and used in later figures. "Median" here refers to the median of the distribution of a variable (e.g. entropy, cooling rate, etc.) at given radius.
}
\begin{center}
\begin{tabular}{ | l | p{70mm} |}
\hline
$K_L$ & Lowest median entropy \\
\hline
$K_H$ & Highest median entropy within $25 ~\text{kpc}$ of the simulation center\\
\hline
$r_L$ & Radius of lowest median entropy \\
\hline
$r_H$ & Radius of highest median entropy  within $25 ~\text{kpc}$ of the simulation center\\
\hline
$r_-$ & Inner radius within which median heating exceeds median cooling \\
\hline
$r_+$ & Outer radius outside of which median heating exceeds median cooling \\
\hline
$t_c(r_x)$ & Median cooling rate at radius $r_x$ \\
\hline
$t_{\text{multi}}$ & Simulation time at which multiphase gas first forms \\
\hline
$r_{\text{multi}}$ & Radius at which multiphase gas first forms \\
\hline
\end{tabular}
\end{center}
\end{table}

\subsection{Condensation of Cold Gas}
\label{sec:formation_of_cold_gas}

\begin{figure}
	\begin{center}
    \includegraphics[width=1\linewidth]{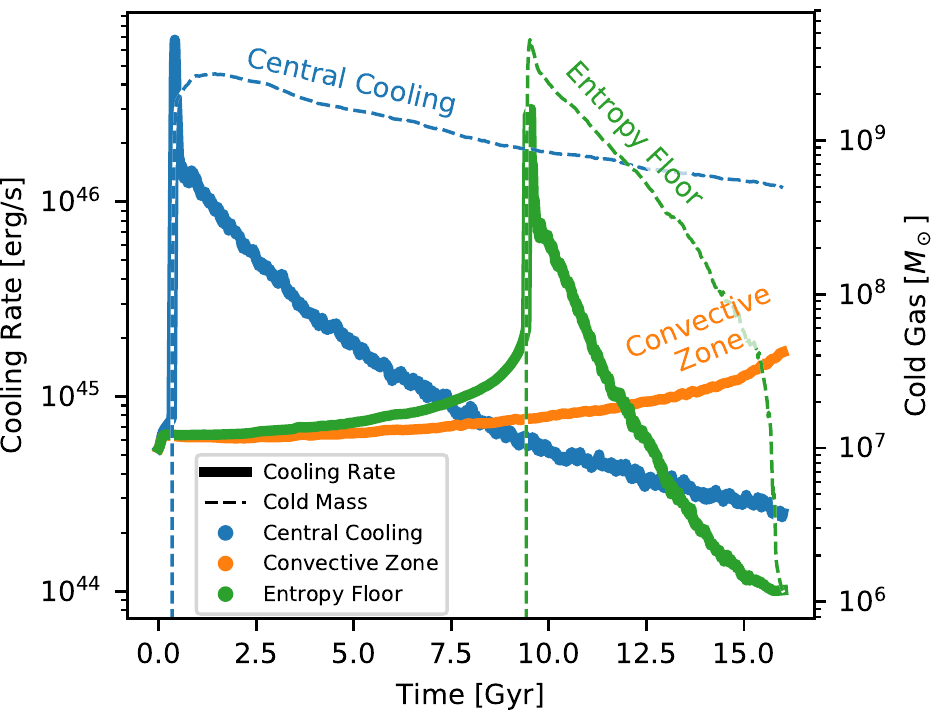}
	\end{center}
	\caption{
    \label{fig:cold_mass_and_cooling}
    Time dependence of total cooling rate (solid lines) and total mass of condensed gas under $3 \times 10^4$ K (dashed lines) for the three simulations shown in Figure \ref{fig:phasePlots}.  The blue points show a simulation with low central heating and excess central cooling ($\alpha=2.0$, $r_s= 8 ~\text{kpc}$, $r_c = 1000 ~\text{kpc}$) that experiences an early cooling catastrophe.  Orange points show a simulation with high central heating ($\alpha=2.6$, $r_s=1 ~\text{kpc}$, $r_c=150 ~\text{kpc}$) that forms a quasi-stable central convective zone.  Green points show a simulation with intermediate central heating ($\alpha=2.6$, $r_s=12 ~\text{kpc}$, $r_c=150 ~\text{kpc}$) that maintains a flat entropy core for almost $10 ~\text{Gyr}$ before undergoing a late cooling catastrophe. In simulations that form a multiphase gas through a cooling catastrophe, the formation of cold gas is preceded by a rise and then a sharp peak in the total cooling rate.  
  }
\end{figure}

Multiphase condensation forms cold gas in many of the simulations, in each case leading to a cooling catastrophe.  Cold gas starts forming near $r_L$, then falls toward the center, displacing buoyantly rising warmer gas. The location of $r_L$ depends on the heating kernel parameters and is related to $r_-$.  

However, when gas at $r_L$ cools enough to transition into the cold phase, it sharply raises the total cooling rate of the halo.  That event immediately boosts the heating rate by the same factor, because our AGN feedback  prescription forces the total heating rate to equal the total cooling rate. This heat is distributed across the halo and is not concentrated on the cooling gas, and thus the AGN feedback does not halt the cooling catastrophe. 

In many cases, rapid heating of lower-density gas during the cooling catastrophe 
produces such great sound speeds
and creates such large discontinuities in the fluid that the simulation becomes infeasible to continue due to the Courant condition.  At that point the heating input greatly exceeds the AGN activity observed in real CC clusters, meaning that the chosen heating kernel has become physically unrealistic.  In simulations that managed to  evolve through this catastrophic event, the heat input leads to drastically elevated entropy in the ambient gas, which slowly reheats the embedded cold gas and prevents more cold gas from forming. After the cooling catastrophe, the core entropy is left much higher than before the catastrophe. Figure \ref{fig:cold_mass_and_cooling} illustrates the timeline of a catastrophe resulting from an increasing cooling rate that leads the formation of cold gas.

\begin{figure*}
	\begin{center}
    \includegraphics[width=1\linewidth]{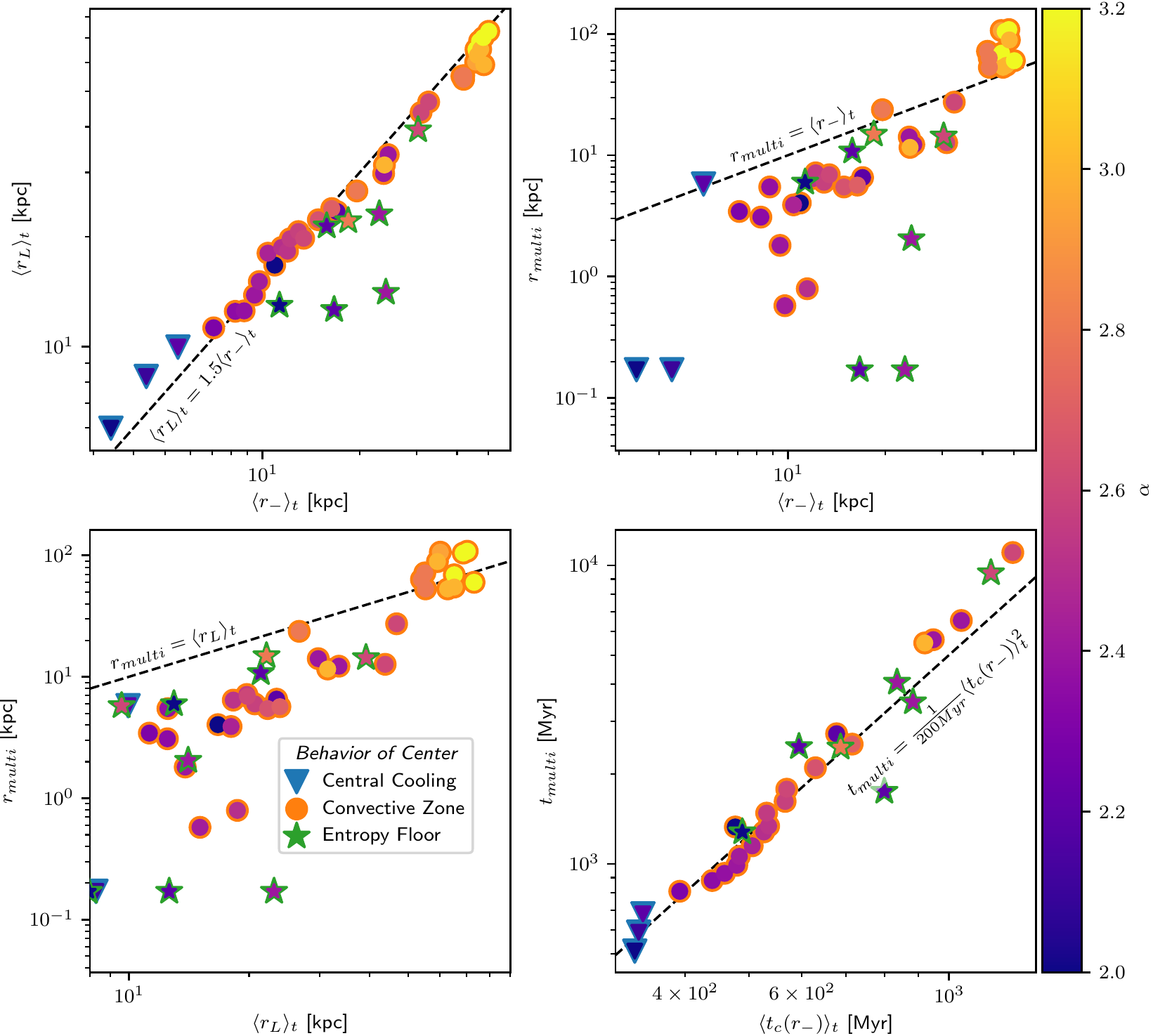}
	\end{center}
	\caption{
    \label{fig:r_minus_effects} Plots of relationships between $r_-$, the radius at which the gas switches from net heating to net cooling, and other features of the simulations.  \textbf{Top left:} Time averaged radius of the minimum of the median entropy profile ($r_L$) versus the time average of $r_-$ up to the formation of a multiphase gas.  (Includes only simulations in which $r_-$ can be defined for at least $50 ~\text{Myr}$.) \textbf{Top right:} Radius at which multiphase gas first forms versus the time averaged $r_-$.  (Includes only simulations in which $r_-$ can be defined for more than one time step.) \textbf{Bottom left:} Radius at which multiphase gas first forms versus the time averaged value of $r_L$ for all simulations. \textbf{Bottom right:} The time required for a simulation to form multiphase gas versus the time averaged value of the cooling time at $r_-$. (Includes only simulations that form multiphase gas and in which $r_-$ can be defined for at least $50 ~\text{Myr}$.)   Shapes in each panel denote the general behavior of the central region of the simulation. Blue highlighted triangles denote Central Cooling simulations, orange highlighted circles denote Central Convective Zone simulations.   Green highlighted stars denote Entropy Floor simulations. Colors show the heating kernel parameter $\alpha$, with greater $\alpha$ generally corresponding to heating that is more centrally concentrated. }
\end{figure*}

Our simulation set generally demonstrates that the radii $r_{\text{multi}}$ and $r_L$ are both related to $r_-$.  Figure \ref{fig:r_minus_effects} shows the relationships among the values of those three radii. We average these quantites over time from the simulation outputs, which have $10 ~\text{Myr}$ frequency, in order to produce one data point per heating kernel.
Larger $\left < r_- \right >$ corresponded to a larger $\left < r_L\right >$, as shown in top right panel, meaning that the radius of lowest entropy corresponds to the inner radius inside of which heating exceeds cooling.  The top right panel shows that larger $\left <r_-\right >$ corresponds to larger $r_{\text{multi}}$, meaning that the radius of lowest entropy corresponds to the inner radius inside of which heating exceeds cooling roughly determines where cold gas first forms. 
In the bottom left panel, $\left <r_L\right >$ also corresponds to larger $r_{\text{multi}}$, showing that multiphase gas typically first forms around the entropy minimum.
The relationship between $r_-$ and the formation of cold gas is most apparent in the plot of $t_c \left (r_- \right)$ versus $t_{\text{multi}}$ in the bottom right panel. When $r_-$ is larger, so that cooling first exceeds heating at a larger radius, the cooling time at $r_-$ is longer, which leads to cold gas forming later in the simulation.  The timescale on which cold gas forms is closely tied to the cooling time of this gas.  Interestingly, the relationship is non-linear, following
\begin{equation}
    t_{\text{multi}} = \frac
    {\langle  t_c \left (r_{\text{multi}} \right) \rangle^2}
    {200 \text{ Myr} } .
\end{equation}
This result is consistent with previous work by \citet{meece_growth_2015} exploring the condensation of gas in the central ICM of galaxy clusters. \citet{meece_growth_2015} found in thermally balanced ICM simulations with varying initial ratios of cooling time to freefall time that gas with a greater initial ratio remains nearly homogeneous for a larger number of cooling times before condensing into a multiphase gas, suggesting a non-linear relationship between cooling time and the formation of a multiphase medium.

\subsection{Central Heating} 
\label{sec:central_heating} 

\begin{figure*}
	\begin{center}
    \includegraphics[width=1\linewidth]{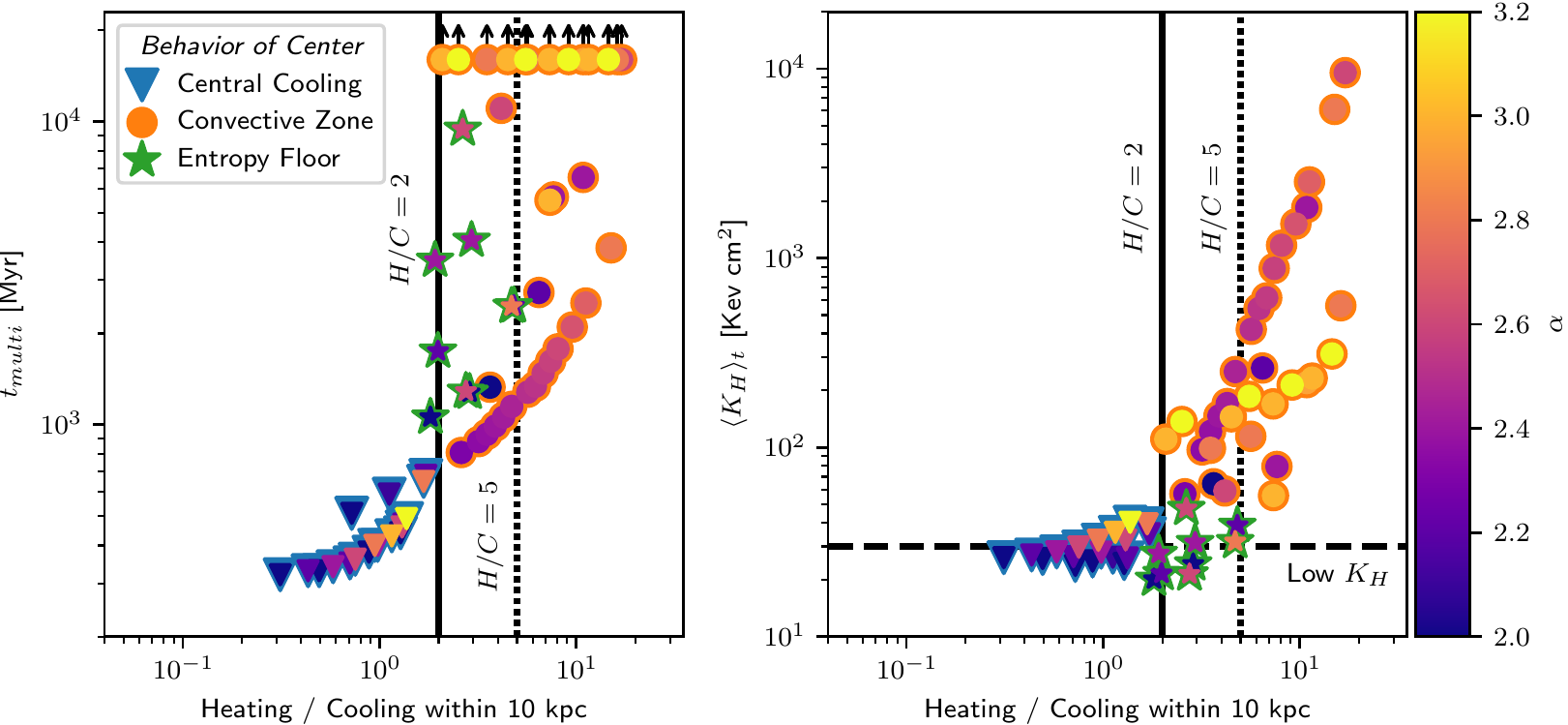}
	\end{center}
	\caption{
    \label{fig:central_heating}
    \textbf{Left:} Time required to form multiphase gas in a simulation versus the ratio of heating to cooling within the inner $10 ~\text{kpc}$ at the first time step.  \textbf{Right:} Maximum of the median entropy within the inner $25 ~\text{kpc}$, versus the ratio of heating to cooling within the inner $10 ~\text{kpc}$ at the first time step. In both panels, a solid line marks a heating to cooling ratio of 2, and a dashed line marks a heating to cooling ratio of 5.   A ratio of at least 2 is required to avoid multiphase condensation within 1~Gyr.  In the right panel, a dashed line marks the maximum central entropy that is observationally expected for a CC cluster.  
  }
\end{figure*}

The heating kernel parameters also affect the central entropy of the cluster halo, in some cases resulting in unreasonably high levels for a CC cluster and in other cases allowing cold gas to quickly condense and collect in the halo center. The central entropy and general behavior of the core is directly related to the amount of heating compared to cooling in the halo center. A certain amount of heating in the center is necessary to offset the central cooling but an excess of heating in the halo center causes central entropies higher than observed in CC clusters. 

To explore this behavior, we track the ratio of the total heating within the inner $10 ~\text{kpc}$ of the halo to the total cooling within the same volume.\footnote{The inner $10 ~\text{kpc}$ volume was chosen to coincide with the region within which the initial entropy profile is nearly flat.  We also tested this analysis using the inner $20 ~\text{kpc}$ volume and found similar results. 
}
Figure \ref{fig:central_heating} shows $t_{\text{multi}}$ and the time average of $K_H$ versus the initial central heating to cooling ratio. A ratio of heating-to-cooling of approximately two is needed to maintain quasi-stability for any significant amount of time, while a ratio greater than five always leads to high central entropies. Inside this range of ratios of heating to cooling, different heating kernels produce all three categories of central entropy behaviors. 

When the integrated heating in the inner region is less than twice the cooling in the same region, a cooling catastrophe happens  within $1 ~\text{Gyr}$. For simulations with less heating than cooling in the central region, cooling quickly causes the central entropy profile to approximate a power law down to the halo center. Cooling gas then flows down the entropy gradient, collecting in the center, and forming multiphase gas. In simulations with average heating one to two times the average cooling rate in the center, density inhomogeneities in the gas allow cooling to exceed heating in some locations.  As the cooling of that gas increases, the total heating rate rises but is insufficient to counter the localized increase in cooling, thus leading a runaway cooling catastrophe. Additionally, as central entropy falls and density increases in the lead up to the catastrophe, central pressure increases and compresses clumps of cooling gas. This further accelerates their cooling during the runaway catastrophe. With simulations having heating-to-cooling ratios above two in the center region, the central cooling is more successfully countered so that the formation of multiphase gas happens on a longer timescale connected to $t_c(r_L)$ and $t_c(r_-)$, as discussed in Section \ref{sec:formation_of_cold_gas}. The left plot in Figure \ref{fig:central_heating} also shows this distinction in behavior.

When central heating rates are more than two times greater than the cooling rate, excess heating leads to central entropies that are higher than what is observed for CC clusters. The right plot in Figure \ref{fig:central_heating} shows the relationship between the ratio of central heating to cooling and the maximum entropy in the central region averaged over time. Some simulations with  two to five times heating to cooling in the center stay under the typical $30 ~\text{keV cm}^2$ specific entropy for CC clusters, but all of the simulations with heating-to-cooling ratios of greater than five produce unrealistically high entropies. With values of $K_H$ above the $30 ~\text{keV cm}^2$ specific entropy where the isentropic entropy profile changes into  power law, these simulations form an inverse convective zone where hot gas collects in the halo center and cold gas collects at $r_L$ at intermediate radii.

\section{Discussion}
\label{sec:discussion}

\subsection{No Adequate Heating Kernel}
\label{sec:no_adequate_heating_kernel}
\begin{figure*}
	\begin{center}
    \includegraphics[width=1\linewidth]{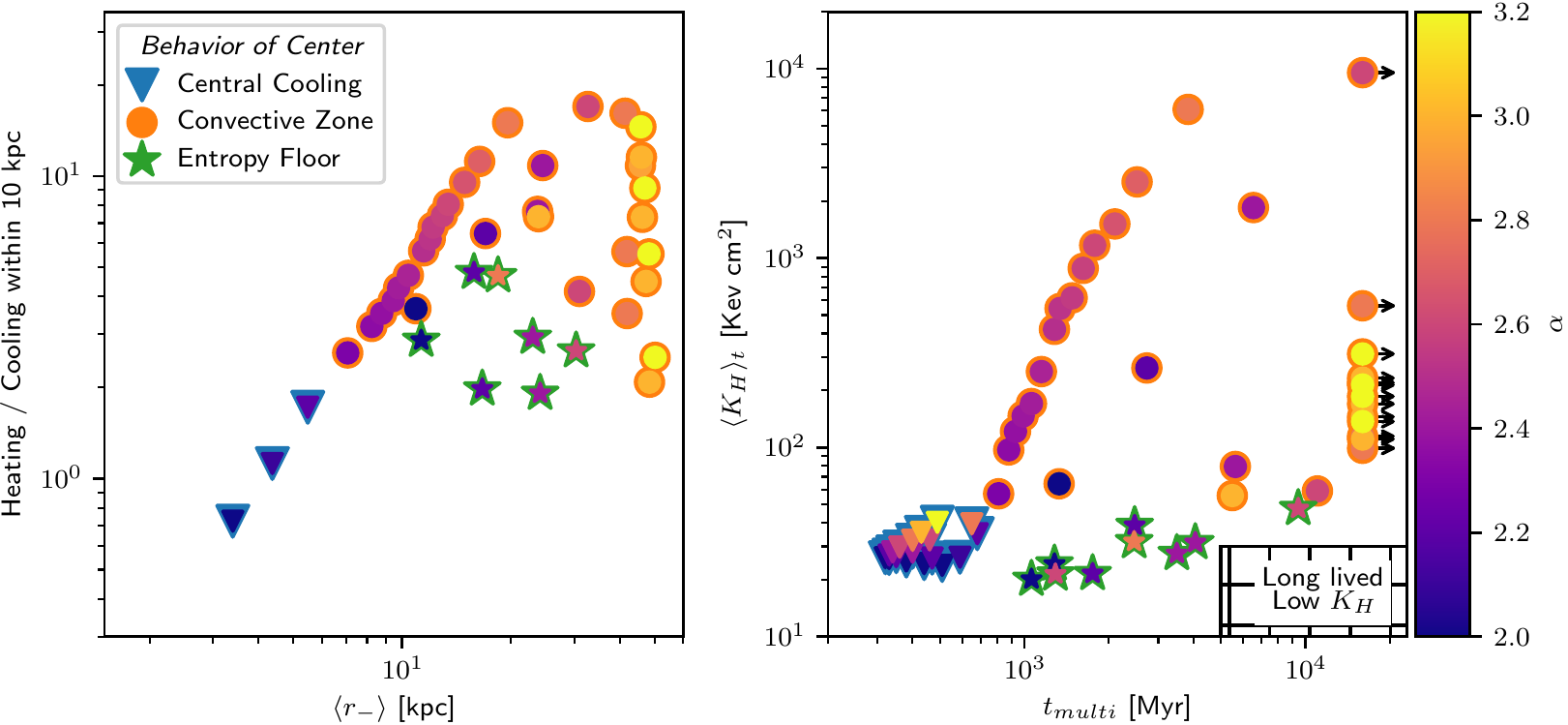}
	\end{center}	
	\caption{
    \label{fig:simulation_adequacy} \textbf{Left:} Relationships between the initial ratio of heating to cooling averaged over the inner $10 ~\text{kpc}$ and the time-averaged radius $\langle r_{-} \rangle$ beyond which cooling begins to dominate over heating.  Only those simulations in which $r_{-}$ can be defined for at least $50 ~\text{Myr}$ are included. 
    The box in the lower right shows hypothetical simulations with an average $r_-$ over $30 ~\text{kpc}$ and an inner heating to cooling ratio under five. 
    \textbf{Right:}  Relationships  between the time average of $K_H$ (the maximum level of the median entropy profile within the inner $25 ~\text{kpc}$) and the time $t_{\rm multi}$ until multiphase gas forms in the simulation.  The plot includes all simulations, assigning $t_{\text{multi}}=16 ~\text{Gyr}$ to simulations that do not form cold gas by that time. An empty box in the lower right corner indicates where points representing heating kernels satisfying adequacy criteria would fall, by persisting for more than $5 ~\text{Gyr}$ before forming multiphase gas while maintaining a maximum entropy level < $30 ~\text{keV cm}^{-2}$ within 25 kpc.  However, no heating kernel we tested satisfies those those criteria.
  }
\end{figure*}

None of the 91 heating kernels we simulated meet all three of the adequacy criteria specified in Section \ref{sec:introduction}.   The failure modes we observe in the simulations can be discussed in terms of the same behavioral categories listed in Section \ref{sec:categorization_of_simulations} for the central entropy profile:
\begin{enumerate}
\item \textbf{Central Cooling.} Heating kernels with low central heating fail to meet our first criterion by producing a cooling catastrophe within $\sim 1$ Gyr that radically changed the structure of the ambient medium.   
\item \textbf{Central Convective Zone.}  Heating kernels with high central heating produces central convective zones that fail to meet our second criterion by producing central entropy levels greatly exceeding those observed among typical CC clusters.  Some of the simulations in this group also fail our longevity criterion because the heating kernel is unable to prevent an early cooling catastrophe due to insufficient heating at intermediate radii.  
\item \textbf{Central Entropy Floor.}  The heating kernels closest to being adequate, according to our criteria, were those with intermediate central heating that exceeds central cooling, but not by a large factor. Those simulations maintain a quasi-stable entropy floor and prevents cooling catastrophe for billions of years.  However, the central entropy profiles of those simulations, while lower than those in the previous category, were still elevated compared to observed CC clusters and thus do not meet our second criterion. Lowering the central heating rates in an attempt to bring their entropy profiles more in line with observation also causes cold gas to form much more quickly.  The simulation that provides results closest to a realistic cluster (with kernel parameters $r_s = 12 ~\text{kpc}$, $r_l = ~\text{kpc}$, and $\alpha=2.4$) maintains a flat entropy core of $30 ~\text{keV cm}^{2}$ and lasts for just under $4 ~\text{Gyr}$, which may be sufficiently long to maintain a CC cluster between external heating events. 
\end{enumerate} 

No heating kernel we tested is able to maintain a low entropy floor close to observations of CC clusters for longer than $4 ~\text{Gyr}$. Figure \ref{fig:simulation_adequacy} summarizes the failure modes of the heating kernels probed in this study. The right panel shows $K_H$ versus $t_{\text{multi}}$, a measure of the longevity of the simulation before a cooling catastrophe strongly altered it. Some simulations prevent a multiphase cooling catastrophe for many Gyr while others maintain low central entropy, but no heating kernel accomplished both aims. The left panel shows the ratio of central heating to cooling versus $r_-$, the two parameters that most strongly influenced the central entropy and longevity, respectively.

\subsection{Robustness of Feedback Algorithm}
\label{sec:robustness_of_heating_prescription}

The ultimate obstacle to finding an adequate thermal heating kernel is the difficulty of preventing gas in the halo center from overcooling while still maintaining a reasonably low entropy profile.   In order to prevent a cooling catastrophe, central heating must be sufficient to raise the median entropy profile enough to keep the lowest-entropy gas from undergoing runaway cooling.   Our simulations show that an integrated central heating rate within the inner $10 \text{ kpc}$ that is approximately two times the cooling rate in that same region is necessary.   Otherwise, too large a proportion of the gas within the central region ends up with cooling exceeding heating, causing a rapid increase in the total radiative cooling rate.  

The consequences of that rapid rise in cooling are dramatic, because the total heating rate is set equal to the radiative cooling rate and rises just as rapidly. However, that heat input is distributed more evenly across a large volume and cannot counteract radiative cooling of localized dense gas clumps.  As a result, the ambient pressure sharply rises, compressing the dense clumps of low-entropy gas, causing both radiative cooling and the matching heating rate to increase.  That coupling therefore causes the cooling/heating rate to spike to unphysically high levels during a cooling catastrophe (see Figure \ref{fig:cold_mass_and_cooling}).  Central internal energies and velocities then rapidly rise and create discontinuities in the fluid. Due to the Courant condition, the time steps sometimes became too small to continue evolving the simulations.  In other cases, those discontinuities lead to negative densities and/orz internal energies in the hydro solver, ultimately ending the simulation.  

In reality, CC clusters can form cold gas (as is evident from observed star formation rates ranging from $1$ to $100 \text{ M}_\odot$ per year), and so a physically accurate model should accommodate the formation of moderate amounts of cold gas.   However, a heating kernel that immediately responds by injecting compensating thermal energy with a fixed spatial distribution appears unable to accommodate multiphase condensation without causing excessive heating.

\subsection{Comparison to Observations}
\label{sec:comparison_to_observations}

\begin{figure}
	\begin{center}
		\includegraphics[width=1\linewidth]{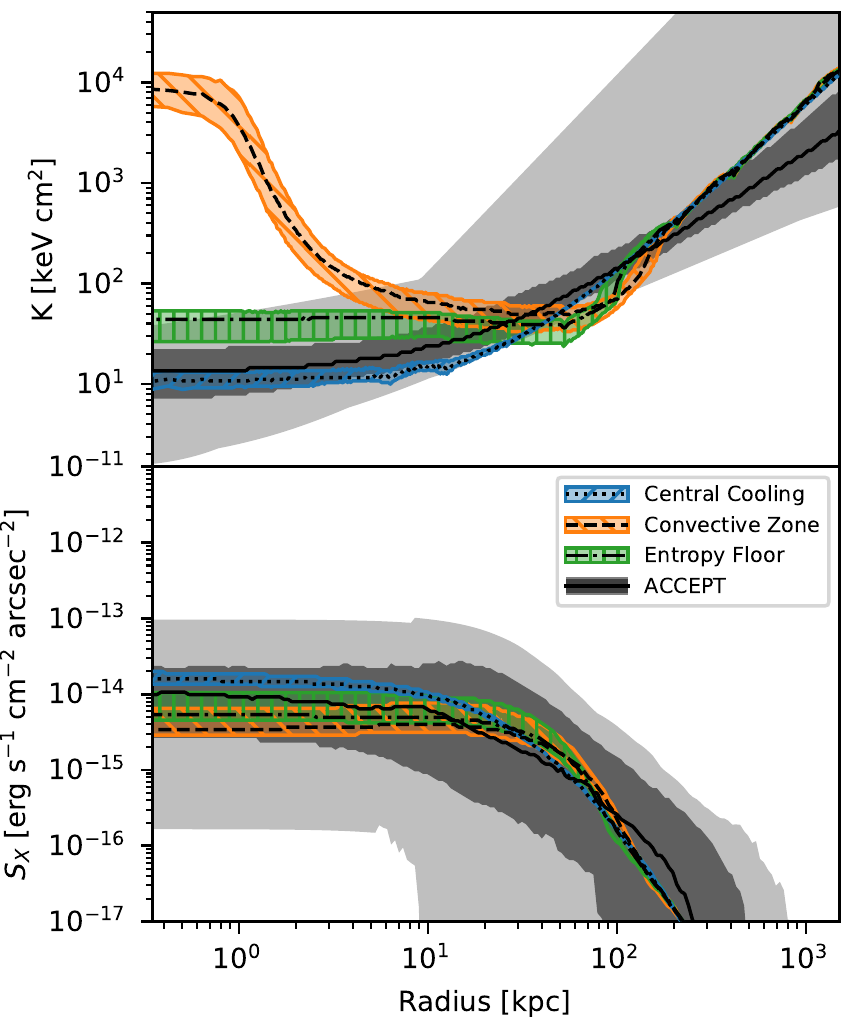}
	\end{center}
	\caption{
    \label{fig:radial_intervals}
    \textbf{Top:} Time-averaged median entropy profiles of the simulated cluster halos in Figure \ref{fig:phasePlots}.   The dotted line shows the simulation with low central heating ( $\alpha=2.0$, $r_s= 8 ~\text{kpc}$, $r_c = 1000 ~\text{kpc}$), and the blue shaded region around it shows the $1\sigma$ dispersion of its median profile over time.  The dashed line shows the simulation with high central heating ($\alpha=2.6$, $r_s=1 ~\text{kpc}$, $r_c=150 ~\text{kpc}$), and the orange shaded region around it shows its $1\sigma$ dispersion.  The dot-dashed line shows the simulation with intermediate central heating ($\alpha=2.6$, $r_s=12 ~\text{kpc}$, $r_c=150 ~\text{kpc}$), and the green shaded region around it shows its $1\sigma$ dispersion.  In each case, entropy is weighted by the x-ray luminosity in the $0.5$--$2.0 ~\text{keV}$ band, to mimic data obtainable with \textit{Chandra}.  The median, $1\sigma$ interval, and full extent of the entropy profiles of clusters with less than $30 ~\text{ keV cm}^2$ from ACCEPT are shown in grayscale, using the broken power law fits from \cite{cavagnolo_intracluster_2009} for the entropy profiles. \textbf{Bottom:} X-ray surface brightness in the $0.5$--$2.0 ~\text{keV}$ band for the same simulated halos, with shaded regions showing the $1\sigma$ dispersion and black lines showing the median. The median, $1\sigma$ interval, and full extent of the entropy profiles of CC clusters from ACCEPT are shown in grayscale, using surface brightness profiles derived from electron density and temperature profiles.
    }
\end{figure}
Figure \ref{fig:radial_intervals} shows the time-averaged median entropy profile and projected X-ray surface brightness profile, along with the $1\sigma$ dispersion in the median profiles.   It also shows the median entropy profile of observed CC clusters in the ACCEPT dataset \citep{cavagnolo_intracluster_2009}, along with the $1\sigma$ dispersion and the full range.   The dispersion in the simulated profiles is computed in radial bins over the lifetime of each simulation up until the formation of cold gas or the end of the simulation.  The dispersion in the ACCEPT data is
generated from a table of power-law fits to the entropy profiles. Only CC clusters from ACCEPT with $K_0 < 30 \text{ keV cm}^2$ are used.

No quasi-stable simulation maintains a central entropy close to the majority of the CC clusters in the ACCEPT dataset. Heating kernels that keep low entropies within the range of the ACCEPT CC clusters are not steady for more than $1 ~\text{Gyr}$, and all experience central cooling catastrophes. Heating kernels that form central convective regions have higher central entropies than the ACCEPT CC clusters. Simulations that form a central entropy floor have lower entropies than the central convective zone simulations and are steady for longer periods than the low central heating kernels, but still have higher central entropies than the majority of observed CC clusters in the ACCEPT dataset.

The differences among the X-ray surface brightness profiles are more subdued, with more centralized feedback corresponding to a lower central surface brightness. The median central surface brightness of the simulation shown here with a central catastrophe is within an order of magnitude of the simulations that form a convective zone. Additionally, the surface brightness profiles from the simulations fall inside the $1\sigma$ interval of the CC clusters from ACCEPT.

\subsection{Comparison to Other Simulations}
\label{sec:comparison_to_other_simulations}

Thermal regulation of galaxy clusters by AGN jets has been studied previously through numerical simulation using many different models of AGN feedback. These approaches include 
injection of buoyant bubbles \citep{fixed-bruggen_simulations_2003,hillel_heating_2016}, 
magnetic fields \citep{li_modeling_2006,nakamura_structure_2006,nakamura_stability_2007,huarte-espinosa_structure_2012}, 
kinetic jets \citep{wu_rhapsody-g_2015,martizzi_rhapsody-g_2016,hahn_rhapsody-g_2017,meece_triggering_2017}, 
stochastic momentum feedback \citep{weinberger_simulating_2017,nelson_first_2019},  
cosmic rays \citep{fixed-jubelgas_cosmic_2008,butsky_role_2018}, and 
turbulent heating \citep{gaspari_mechanical_2012,zhuravleva_turbulent_2014,banerjee_turbulence_2014}, either explicitly or implicitly driven by the central SMBH.
Some simulations have also used purely thermal feedback models like the model used in this work, to which we can compare.

\citet{meece_triggering_2017}, the predecessor to this work, tested a AGN feedback model consisting of a precessing bipolar jet that injected kinetic and thermal energy. 
They tested different fractions of AGN feedback going into thermal heating versus the kinetic jet. For triggering the feedback they tested three different models: a cold gas triggering model from \citet{li_modeling_2014}, a boosted Bondi-like triggering, and a Booth and Schaye accretion model \citep{booth_cosmological_2009}. Like this work, \citet{meece_triggering_2017} found that AGN models with purely thermal feedback led to an overabundance of cold gas in the simulation core. However, their thermal feedback was limited to a small region around the AGN, less than $1 ~\text{kpc}$ in diameter. 
In their simulations, hot bubbles inflated via AGN heating at the cluster center buoyantly rose a short distance out of the center to $10-30 \text{ kpc}$ and created a flatter entropy profile that was unstable to multiphase condensation
and therefore failed to suppress large accumulations of multiphase gas. Many of the heating kernels tested in this paper rectify the problem of overly centralized heating but result in elevated core entropy beyond what is reasonable for a CC cluster. 

The \RhapsodyG simulations of galaxy clusters explored cosmological zoom-in simulations with star formation and feedback (SFF) and supermassive black hole (SMBH) formation and feedback, using the \Ramses Eulerian AMR code \citep{wu_rhapsody-g_2015,teyssier_cosmological_2002}. In their AGN feedback prescription, mass accreted onto the SMBH following a density-boosted Bondi-Hoyle accretion rate~\citep{booth_cosmological_2009}.  Thermal energy was deposited into a small radius around the SMBH \citep{martizzi_rhapsody-g_2016}. 
Compared to CC cluster entropy profiles from the ACCEPT catalogue, CC clusters in the \RhapsodyG had lower central entropies, showing overcooling in the inner tens of kpc \citep{hahn_rhapsody-g_2017}.

\citet{tremmel_romulus_2017} presented the \Romulus galaxy simulations using the \CHANGA smoothed particle hydrodynamics code and includes SMBH feedback and SFF models tuned to observations. Their SMBH feedback model had two free parameters: (1) the efficiency of the accretion rate  onto the SMBH and (2) the gas coupling efficiency $\epsilon_c$. These parameters were calibrated to produce galaxies with observed values of the stellar-mass to halo ratio, HI gas fraction as a function of stellar mass, galaxy specific angular momentum versus stellar mass, and the SMBH to stellar mass relation. Their simulations used a thermal-only feedback model that deposited feedback energy into the 32 gas particles nearest to the SMBH. Mass accretion was governed by a modified Bondi accretion rate. Gas cooling was suppressed when heated by the SMBH for a time step equal to the time step of the SMBH. This allowed energy to escape away from the SMBH, although it may not be physically realistic. This feedback model produced galaxies with regulated SFF compared to observation.

In the follow-up paper \citet{tremmel_introducing_2019} on the cosmological \RomulusC simulations, the same SFF and SMBH feedback models were used in a zoom-in simulation of a single halo. In an isolated halo, purely thermal feedback from the SMBH led to a conic structure with a highly collimated jet-like outflow. The outflows evolved over time, changing in shape and direction with the angular momentum of the gas near the SMBH. Energy was carried out to large radii through the outflows, which suppressed cooling at large radii. Star formation rates were regulated and matched observed rates in clusters. Additionally, the entropy profile of the clusters was within the range of observed profiles in CC clusters. Although the outflows were not explicitly introduced by their feedback prescription, their ability to transport AGN feedback energy tens of kiloparsecs from the center without inverting the large-scale entropy profile and overstimulating thermal instability is the key to proper thermal regulation of their simulated CC cluster.

\subsection{Implications}
\label{sec:implications}

Since the heating kernels explored here failed to produce quasi-stable CC clusters with realistic entropy profiles, extrapolations to real CC clusters may not be accurate. However, a few lessons can be drawn from these simulations:
\begin{itemize}

\item In the context of purely thermal AGN feedback, feedback that is highly centrally concentrated and tied directly to the global radiative cooling rate produces cores with entropy levels that greatly exceed those of observed CC clusters and in some cases are physically unreasonable.

\item When the total heating rate is directly tied to the total cooling rate in the halo, rapid cooling of gas into cold clumps causes the heating rate to reach unphysically high levels. In comparison, in simulations using Bondi accretion or cold gas accretion such as in \citet{meece_triggering_2017} AGN feedback increases more gradually with the formation of cold gas, allowing feedback energy output to tune itself to physically reasonable values.

\item The heating kernels considered here, in which heating per unit volume had a fixed radial distribution, were unable to maintain thermal stability of the cluster halo. In cases where a cold clump of gas formed, the purely thermal AGN feedback was insufficient to disrupt the clump without injecting unphysically high amounts of energy. The thermal heating in these simulations was unable to reproduce the effects caused by kinetic outflows from AGN jets such as in \citet{meece_triggering_2017}.

\end{itemize}

A spherically symmetric heating kernel for purely thermal feedback that satisfies all of our criteria may exist but would need to have different parameters than are explored here. Such an idealized heating kernel would  be useful to efficiently include AGN feedback in cosmological simulations.

\subsection{Other Models Investigated}
\label{sec:other_models_investigated}

In search of a satisfactory heating kernel, we investigated several extensions to the spherically symmetric ones described in Section \ref{sec:methodology}. First, we applied a polar angle dependence of $\cos^2 \theta$ to mimic the conical distribution of heat from a kinetic jet. Total heating remained linked to total cooling. However, decreased heating near the equatorial plane leads to cold gas forming several tens of Myr sooner than for the corresponding spherical kernel and did not change the general behavior of the cooling catastrophe. Next, we tried a model in which cold gas was removed from the center of the simulation as it formed, to decrease the central density, potentially avoid fluid discontinuities in the fluid solver, and allow robust simulations with the formation of cold gas. However, explosive heat input triggered by the formation of cold gas still causes the hydrodynamics solver to fail. We also tested equating total heating to total cooling of only the warm gas, testing separately temperature thresholds of $10^{6.5} ~\text{K}$ and $10^7 ~\text{K}$, to exclude the rapid cooling of cold gas and avoid explosive AGN feedback. However, this filtering of cold gas in the calculation of the heating rate leads to more cold gas forming and the leftover warm gas having an elevated central entropy. In some cases the heat input is still great enough to halt the simulation because of the Courant condition. Lastly, we tried smoothing out the rise in AGN heating by setting the total feedback to the average of the cooling rate over the last $50 ~\text{Myr}$, in essence implementing a temporal kernel as well as a spatial kernel.   However, this approach also leads to high rates of formation of cold gas due to the delayed heating response, as well as an eventual spike in AGN heating since the cooling catastrophe ultimately is not counteracted.

\subsection{Future Models}
\label{sec:future_models}

There remain conceivable modifications to this heating kernel approach that we did not investigate, but which could produce more physically realistic CC clusters.  For example, total heating could be capped at a physically reasonable value to avoid the overheating that coincides with the formation of cold gas. Additionally, we could investigate a radially piecewise conic feedback kernel in which AGN heating is spherically symmetric at small radii and conical at large radii.  Another alternative would be a kernel with a spatial distribution that depends on the total heat input, adjusting to spikes in heating/cooling by distributing increased heating over a larger volume, as would happen with an increase in total jet power.

\section{Summary}
\label{sec:summary}

We have presented simulation results for simplified models of AGN feedback using heating kernels for purely thermal feedback. In those kernels, heat input has a spatial dependence following a radial power law $\dot e \propto r^{-\alpha}$ having a smoothing length $r_s$ at small radii, an exponential cutoff radius $r_c$ at large radii, and a total heating rate set equal to the total
cooling rate measured within the cluster halo. This approach differs from previous simulations approximating feedback rates using Bondi and cold gas accretion models, which can temper the feedback response but are computationally more expensive. Our intention was to identify a heating kernel that would be both computationally inexpensive and able to maintain the hot atmosphere of a galaxy cluster in realistic quasi-steady state. 

All of the heating kernels we tested failed to maintain a quasi-steady state with an entropy profile consistent with those observed among cool-core clusters (see Figures \ref{fig:phasePlots} and \ref{fig:simulation_adequacy}). We compared entropy profiles from our simulations to observational data from the ACCEPT dataset. Some simulations exhibit small to large central peaks in entropy that differ significantly from the entropy profiles seen in the ACCEPT sample. The central entropy peaks are most pronounced in simulations with highly centralized feedback. Simplified AGN models with overly centralized thermal heating therefore do not produce realistic entropy profiles.

A few lessons can be drawn from this work. Thermalization of AGN feedback energy must occur over a large region in order for the entropy profiles of simulated clusters to agree with those of observed cool-core clusters. However, it is difficult to distribute thermal feedback over a large region while also preventing a cooling catastrophe.  Also, requiring total heating to equal total cooling becomes particularly problematic near the onset of a cooling catastrophe, because the increased cooling rate during the formation of large clumps of cold gas raises the heating rate to very high levels.

No configuration of purely thermal feedback explored here achieved thermal stability nor prevented a run away collapse into a cold clump, in contrast to simulations that introduce feedback energy in the form of kinetic jets.  A heating kernel for purely thermal AGN feedback that produces realistic CC clusters may still exist but would need to significantly differ from the kernels we tested. Such a heating kernel that functions as an accurate and efficient proxy for more complex AGN feedback physics would allow larger cosmological simulations without increasing resolution.

\acknowledgments
We thank Philipp Grete and Deovrat Prasad for useful discussions. We also thank Megan Donahue, Dana Koeppe, and Rachel Frisbie for assistance with the ACCEPT database.

This project has been supported by NASA through Astrophysics Theory Program
grant \#NNX15AP39G and Hubble Theory Grant HST-AR-13261.01-A, and by the
NSF through grant AST-1514700.  The simulations were run on the NASA
Pleiades supercomputer through allocation SMD-16-7720 and at the Michigan State University High Performance Computing Center (operated by the Institute for Cyber-Enabled Research).  \texttt{Enzo} and
\texttt{yt} are developed by a large number of independent researchers from
numerous institutions around the world. Their commitment to open science
has helped make this work possible.

\bibliographystyle{aasjournal}

\end{document}